\documentclass[twocolumn,english,aps,prl,groupedaddress,superscriptaddress]{revtex4}
\usepackage{graphicx,epsfig,units}
\usepackage{xcolor} 
\usepackage{soul} 
\usepackage{amsmath,amsfonts,mathrsfs,amsbsy,bm,babel}

\begin{document}

\title{Robust antiferromagnetic spin waves across the metal-insulator transition in hole-doped BaMn$_{2}$As$_{2}$ }

\author{M.~Ramazanoglu}
\affiliation{Physics Engineering Department, Istanbul Technical University, 34469, Maslak, Istanbul, Turkey}
\affiliation{T{\"{u}}bitak B{\.I}DEP 2232 Fellow, Ankara, 06100, Turkey}

\author{A.~Sapkota}
\affiliation{Ames Laboratory, Ames, IA, 50011, USA}
\affiliation{Department of Physics and Astronomy, Iowa State University, Ames, IA, 50011, USA}

\author{Abhishek~Pandey}
\altaffiliation[Present address: ] {Department of Physics and Astronomy, Texas A\&M University, College Station, TX 77843, USA}
\affiliation{Ames Laboratory, Ames, IA, 50011, USA}
\affiliation{Department of Physics and Astronomy, Iowa State University, Ames, IA, 50011, USA}

\author{J.~Lamsal}
\affiliation{Ames Laboratory, Ames, IA, 50011, USA}
\affiliation{Department of Physics and Astronomy, Iowa State University, Ames, IA, 50011, USA}

\author{D.~L.~Abernathy}
\affiliation{Quantum Condensed Matter Division, Oak Ridge National Laboratory, Oak Ridge, TN, 37831, USA}

\author{J.~L.~Niedziela}
\affiliation{Quantum Condensed Matter Division, Oak Ridge National Laboratory, Oak Ridge, TN, 37831, USA}

\author{M.~B.~Stone}
\affiliation{Quantum Condensed Matter Division, Oak Ridge National Laboratory, Oak Ridge, TN, 37831, USA}

\author{A.~Kreyssig}
\affiliation{Ames Laboratory, Ames, IA, 50011, USA}
\affiliation{Department of Physics and Astronomy, Iowa State University, Ames, IA, 50011, USA}

\author{A.~I.~Goldman}
\affiliation{Ames Laboratory, Ames, IA, 50011, USA}
\affiliation{Department of Physics and Astronomy, Iowa State University, Ames, IA, 50011, USA}

\author{D.~C.~Johnston}
\affiliation{Ames Laboratory, Ames, IA, 50011, USA}
\affiliation{Department of Physics and Astronomy, Iowa State University, Ames, IA, 50011, USA}

\author{R.~J.~McQueeney}
\affiliation{Ames Laboratory, Ames, IA, 50011, USA}
\affiliation{Department of Physics and Astronomy, Iowa State University, Ames, IA, 50011, USA}

\begin{abstract}
BaMn$_{2}$As$_{2}$ is an antiferromagnetic insulator where a metal-insulator transition occurs with hole doping via the substitution of Ba with K.  The metal-insulator transition causes only a small suppression of the N\'eel temperature ($T_\mathrm{N}$) and the ordered moment, suggesting that doped holes interact weakly with the Mn spin system. Powder inelastic neutron scattering measurements were performed on three different powder samples of Ba$_{1-x}$K$_{x}$Mn$_{2}$As$_{2}$ with $x=$0, 0.125 and 0.25 to study the effect of hole doping and metallization on the spin dynamics of these compounds.  We compare the neutron intensities to a linear spin wave theory approximation to the $J_{1}-J_{2}-J_{c}$ Heisenberg model. Hole doping is found to introduce only minor modifications to the exchange energies and spin gap.  The changes observed in the exchange constants are consistent with the small drop of $T_\mathrm{N}$ with doping.  
\end{abstract}

\date{February 15, 2017}
\maketitle

\section{Introduction}
The parent compounds of unconventional superconductors are typically antiferromagnetic (AFM), although they may be initially metals (as in the iron pnictides \cite{Kamihara}) or Mott insulators (as in the copper oxide superconductors \cite{Imada}).  In either case, chemical substitution is often employed to destabilize the AFM ordered state and give rise to a superconducting ground state over some composition range.  For the iron arsenides, they are already metallic and adding charge carriers serves to modify the Fermi surface and destabilize nesting-driven spin-density wave AFM order, resulting in superconductivity.   In BaFe$_{2}$As$_{2}$, electrons or holes can be added via chemical substitutions such as Ba(Fe$_{1-x}M_x$)$_2$As$_{2}$ with $M=$ Co or Ni \cite{Lee,Huang,Sefat,Li,Kurita,Leithe-Jasper,Kim12} and Ba$_{1-x}$K$_{x}$Fe$_{2}$As$_{2}$ \cite{Rotter}, respectively.   

On the other hand, the effect of chemical substitutions in the copper oxides, such as La$_{2}$CuO$_{4}$, are two-fold.  Since they are insulators, chemical substitutions (such as La$_{2-x}$Sr$_{x}$CuO$_4$) must both metallize the system and disrupt long-range AFM order in order to make the conditions favorable for superconductivity.  In other words, both a metal-insulator transition and suppression of AFM order are necessary for superconductivity to appear in the cuprates.  

In attempting to find some commonality between the arsenides and cuprates, we search for other compounds that bridge these two systems.  One possible system is BaMn$_{2}$As$_{2}$ which shares the same crystal structure as BaFe$_{2}$As$_{2}$.  Similar to cuprates, BaMn$_{2}$As$_{2}$ is a quasi-two-dimensional AFM insulator possessing a square lattice of magnetic moments that order into a G-type AFM (checkerboard) pattern \cite{singh09}.   Also similar to the cuprates, a metal-insulator transition can be induced in BaMn$_{2}$As$_{2}$ by replacing small amounts (a few percent) of Ba with K which effectively adds hole carriers \cite{Bao,pandey12}.  Unlike the cuprates, the moments on Mn are large ($ S \approx$ 2 to 5/2).  In conjunction with large magnetic interactions (with a N\'eel transition temperature of $T_{\mathrm{N}}=$ 625 K) \cite{singh09,singh092,johnston11}, both neutron diffraction \cite{Lamsal} and $^{75}$As nuclear magnetic resonance \cite{yeninas} find that long-range AFM ordering is robust with hole doping, even at K concentrations up to 40\%.

Measurements and calculations of the electronic band structure of BaMn$_2$As$_2$ indicate that significant hybridization exists between Mn and As orbitals \cite{An,mazin, zhang}. Thus, we expect there to be a strong effect of hole doping on magnetic exchange interactions, Mn moment size and AFM order.  This is not borne out, based on the weak suppression of $T_{\mathrm{N}}$ and the ordered magnetic moment.  In addition, it is surprising to find that weak ferromagnetism (FM) appears at higher hole concentrations ($x \geq $ 0.16) which coexists with long-range AFM order \cite{Bao, pandey13, pandey15}.   A strong x-ray magnetic circular dichroism (XMCD) signal was observed at the As $K$-edge which shows that ferromagnetic ordering originates in As 4$p$ conduction bands \cite{ueland15}. In principle, observation of itinerant ferromagnetism in the presence of local moment AFM order suggests that charge transport and antiferromagnetism in Ba$_{1-x}$K$_{x}$Mn$_{2}$As$_{2}$ are largely decoupled.

To further address the connection between hole doping and AFM order in BaMn$_{2}$As$_{2}$, inelastic neutron scattering (INS) was used to probe the AFM spin excitations in
 BaMn$_{2}$As$_{2}$ and two different K-substituted Ba$_{1-x}$K$_{x}$Mn$_{2}$As$_{2}$ polycrystalline samples with $x=$ 0.125 and 0.25.   The INS spectrum was analyzed using an AFM Heisenberg spin-wave model based on $J_{1}-J_{2}-J_{c}$ nearest-neighbor (NN) and next-nearest-neighbor (NNN) interactions between spins. Our results show that the introduction of hole carriers, even at concentrations up to 12.5\% per Mn ion ($x=$ 0.25), has only a minor effect on the spin waves. This supports the idea that doped holes occupy hybridized As bands, similar to a charge-transfer insulator, and weakly affect the magnetic moment on Mn and AFM exchange interactions between Mn spins.

\section{Experimental Results}
Magnetic susceptibility, resistivity, ARPES, and heat capacity measurements show that pure BaMn$_2$As$_2$ is an AFM insulator \cite{johnston11}. Large Mn local moments with a magnitude of approximately 4 $\mu_\mathrm{B}$ are observed in high-temperature susceptibility and neutron diffraction experiments.  A well-defined charge gap of 0.86 eV is measured using optical spectroscopy which attests to the insulating properties \cite{McNally}. The results of these measurements are reproduced by dynamical mean-field theory calculations showing that the parent BaMn$_2$As$_2$ compound is a Mott-Hund insulator.  Weakly K-doped samples show metallic behavior without disturbing the parent AFM state and more highly K-doped compositions show weak FM metallic properties below $T_{\mathrm{C}} \approx$ 100 K while retaining the high N\'eel temperature ($T_\mathrm{N}$ drops to 480 K for $x=$ 0.4) \cite{Bao, pandey12, Lamsal}.

INS measurements were performed on powders of BaMn$_{2}$As$_{2}$ and K-substituted Ba$_{1-x}$K$_{x}$Mn$_{2}$As$_{2}$. BaMn$_2$As$_2$ has a body-centered tetragonal $I4$/$mmm$ structure with lattice parameters $a= $4.15 \AA\ and $c=$ 13.41 \AA.\cite{Lamsal} In addition to the parent compound, two different substituted samples were prepared for this measurement with $x=$ 0.125 and 0.25 (with hole concentrations of $\delta = x/2$ per Mn ion). The powders, each with mass of roughly 7 grams, were prepared by conventional solid-state reaction.  Each sample was packed into a cylindrical Al sample can for neutron scattering measurements.  The ARCS neutron spectrometer at Oak Ridge National Laboratory was used to collect INS profiles with different incident neutron energies ($E_i$) of 30, 74, 144.7 and 315 meV. The time-of-flight data were reduced into energy transfer ($E$) and momentum transfer ($Q$) profiles and data corrections for detector efficiency and the empty aluminum can were performed. The $S$($Q$,$E$) scattering profiles and constant energy/momentum cuts were obtained with Mslice software \cite{Dave}.

\begin{figure}
\includegraphics[width=0.9\linewidth]{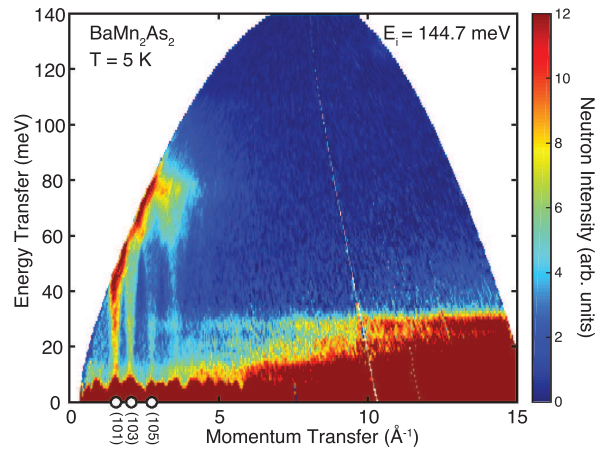}
\caption{\footnotesize Inelastic neutron scattering data from a powder sample of BaMn$_{2}$As$_{2}$ plotted as a function of momentum transfer ($Q$) and energy transfer ($E$) with $E_{i}=$ 144.7 meV.  The large white regions correspond to areas that are not accessible by the experiment due to kinematic restrictions of the scattering process.  The color scale corresponds to the intensity of scattered neutrons obtained after the scattering from an empty aluminum sample can has been subtracted.  Magnetic excitations appear as steeply dispersing modes at low-$Q$ emanating from magnetic Bragg peaks (101), (103), (105), etc., while phonons appear as flat bands below 40 meV whose intensity increases as $Q^2$.}
\label{fig1}
\end{figure}

The results of powder INS measurements conducted on the parent BaMn$_{2}$As$_{2}$ compound at $T=$ 5 K are shown over the accessible energy and momentum transfer ranges for $E_{i}=$ 144.7 meV in Fig.~\ref{fig1}.  Our collaboration has performed similar INS measurements on the parent compound previously and obtained qualitatively similar results \cite{johnston11}, as discussed later.  Even though these unpolarized INS powder intensities include both lattice and magnetic excitations, one can distinguish magnetic and phonon contributions by their different $Q$-dependences. The flat phonon bands exist mainly below 40 meV and their intensities increase as $Q^{2}$.  The steeply dispersive features at low $Q$ are identified as magnetic in origin since their intensity falls off with $Q$ consistent with the Mn$^{2+}$ magnetic form factor.  In addition, the steep dispersions originate from magnetic Bragg peaks of BaMn$_2$As$_2$, such as $Q_{(101)}=$ 1.58 \AA\ and $Q_{(103)}=$ 2.06 \AA, providing additional confirmation of their magnetic character.\cite{Lamsal}

\begin{figure}
\includegraphics[width=1\linewidth]{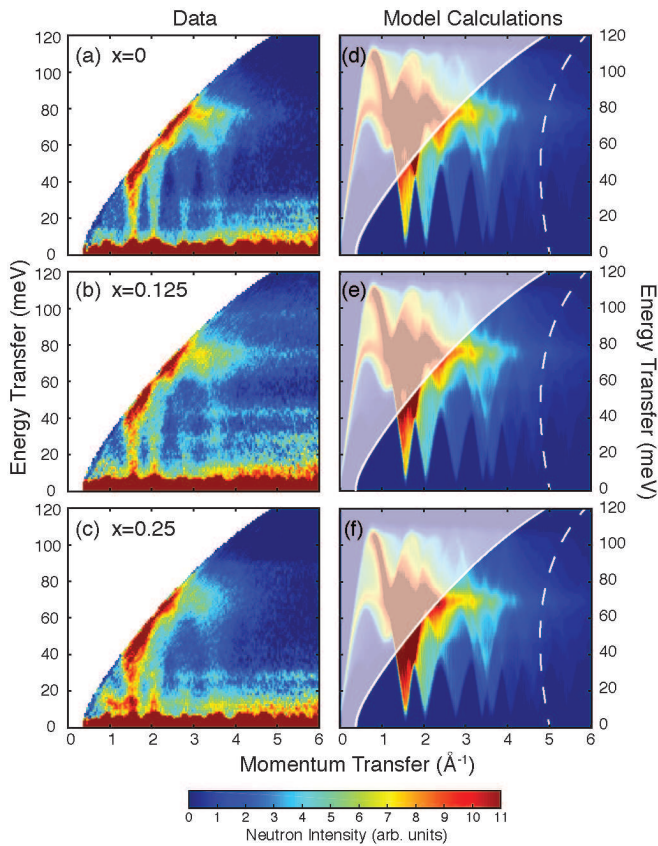}
\caption{\footnotesize Inelastic neutron scattering data measured with $E_{i}$=144.7 meV from powders of BaMn$_{2}$As$_{2}$ in panel (a) and K-doped Ba$_{1-x}$K$_{x}$Mn${_2}$As$_{2}$ in panels (b) $x=$ 0.125 and (c) $x=$ 0.25.  Data panels are drawn over a smaller $Q$ range as compared to Fig.~\ref{fig1}.   Panels (d), (e) and (f) show neutron intensities calculated using linear spin wave theory of the Heisenberg model for corresponding powders using parameters obtained by fitting the data.  Shaded intensities in the upper left of panels (d), (e) and (f) are inaccessible by the experiment due to kinematic constraints defined by the lowest scattering angle of $2\theta_{\mathrm{min}}=2.5$ degrees.  The dashed white line on the right corresponds to a scattering angle of $2\theta=35$ degrees and was used as a cutoff for the summation of the magnetic spectra shown in Fig.~\ref{fig3}.}
\label{fig2}
\end{figure}

Panels (a), (b) and (c) of Fig.~\ref{fig2} focus on the spin wave scattering observed at low $Q$ with $E_{i}=$ 144.7 meV for $x=0$, 0.125 and 0.25, respectively.  The steeply dispersing portions up to $\sim$ 50 meV are acoustic spin waves and the slope of this dispersion in the vicinity of $Q_{(101)}$ is related to $v_{ab}$, the in-plane spin wave velocity.  A rough estimate gives $v_{ab} \approx$ 200 meV \AA.  Contributions of spin waves close to the magnetic Brillouin zone boundary produce broad, featureless bands observed between 50 and 100 meV.  These features of the spin wave spectrum and their doping dependence will be analyzed in detail below.  But, even from the raw data shown here, the main features and dispersions seen in the parent compound are qualitatively similar to the K-doped samples.

\section{Analysis and Modeling}
\subsection{Spin wave spectrum}
The simplest first step is to analyze the energy spectrum of spin waves obtained from averaging INS data over the low $Q$ region (over a scattering angle range from $2\theta=$ 2.5$-$35 degrees, as shown in Fig.~\ref{fig2}.  As is apparent in Figs.~\ref{fig1} and \ref{fig2}, phonon scattering and other background sources contribute to the total scattering intensity at low $Q$ and a determination of these contributions is a crucial step for isolating and fitting the magnetic spectrum properly.   A  discussion of the phonon scattering and other background estimates can be found in the Appendix.  The phonon scattering intensity increases with $Q^{2}$ while the spin wave scattering intensity drops monotonically with $Q$ following the Mn$^{2+}$ form factor.  Therefore, the phonon signal can be estimated from the high $Q$ data where the magnetic signal is very weak or absent. 

\begin{figure}
\includegraphics[width=0.8\linewidth]{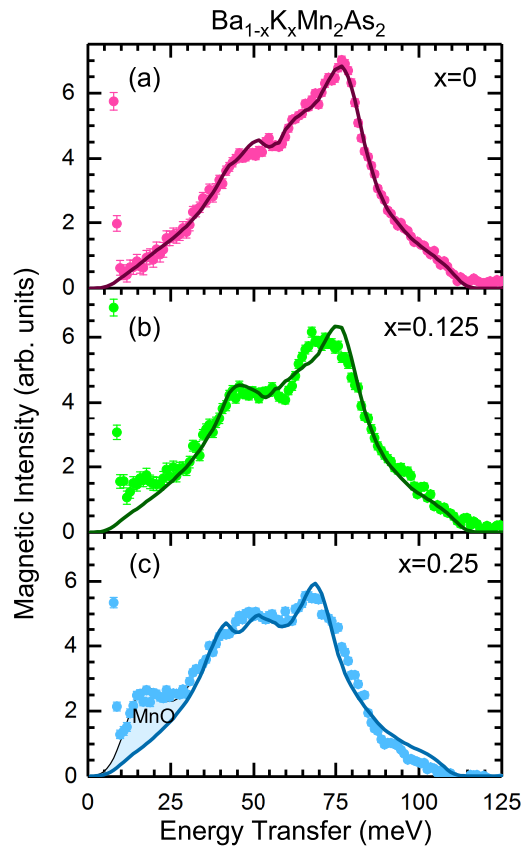}
\caption{\footnotesize Magnetic spectra for Ba$_{1-x}$K$_{x}$Mn$_{2}$As$_{2}$ as a function of energy for (a) $x=0$, (b) $x=0.125$ and (c) $x=0.25$.  Symbols show the neutron data that were obtained with $E_{i}=$ 144.7 meV and summed over a scattering angle range from 2 to 35 degrees.  Subtraction of phonons and other backgrounds are discussed in the Appendix.  Lines correspond to fits obtained using the Heisenberg model described in the text.   For $x=0.25$, the additional magnetic scattering observed at low energies (shaded region) is consistent with antiferromagnetic spin wave scattering from MnO  \cite{kohgi}.}
\label{fig3}
\end{figure}

Closer examination of the phonon scattering (see Appendix) and diffraction data on the elastic line reveal that some impurity phases exist in the K-doped samples.  For example, it is likely that oxide impurities result in some of the high energy phonon background near 80 meV in the $x=$ 0.125 sample.  MnO is one of these oxides and we observe weak magnetic Bragg peaks originating from long-range AFM order of MnO.  We also find weak contributions from impurity MnO spin wave scattering below 25 meV for $x=$ 0.125 and 0.25 samples (see Figs.~\ref{fig2} and \ref{fig3}).  In order to estimate this background contribution, calculations of the spin wave scattering for pure MnO powder were performed on the basis of published Heisenberg exchange constants \cite{kohgi}, as described in the Appendix. 

The magnetic spectra obtained after subtracting background contributions are shown in Fig.~\ref{fig3} for each of the compositions.  The magnetic spectrum for the parent compound is characterized by a peak near 75 meV and an upper cutoff of $\sim$ 115 meV, similar to that published previously \cite{johnston11}.  The K-substituted samples, especially the more heavily substituted $x=$ 0.25 composition, show some deviations from the parent compound, such as a shift of the main spectral peak to lower energies ($\sim$ 70 meV) and some spectral broadening.  The mean energy of the distribution, $\langle E \rangle$, (averaged over an energy range from 30 to 110 meV) changes from 66.0 to 59.2 for $x=$ 0 and 0.25, respectively, in rough correspondence with the decrease in $T_{\mathrm{N}}$ (see Table \ref{tablo1}). These doping-dependent effects can arise from a lowering of the average exchange energy with hole doping, disorder, or from damping effects associated with the introduction of hole carriers into the system.  This will be discussed after performing a more detailed Heisenberg model analysis described below.

\subsection{Spin gap}
\begin{figure}
\includegraphics[width=1\linewidth]{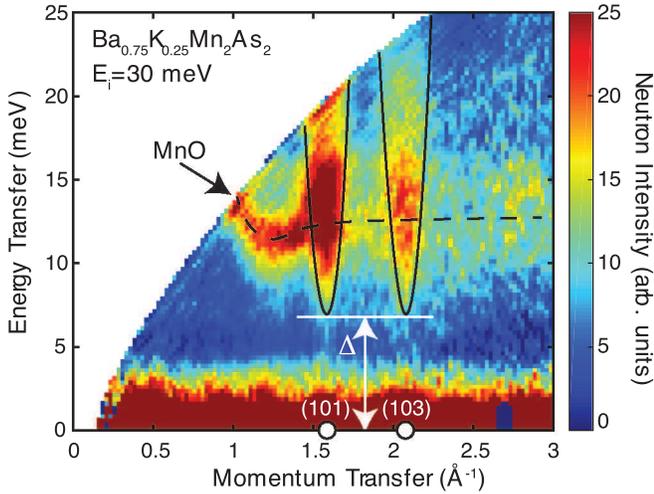}
\caption{\footnotesize $S(Q,E)$ profile of the Ba$_{0.75}$K$_{0.25}$Mn$_2$As$_2$ sample measured with $E_{i}=$ 30 meV. Steeply dispersive excitations from $Q_{(101)}$ and $Q_{(103)}$ Bragg peaks are highlighted by the solid black outlines.  It is evident that a well-defined energy gap ($\Delta$) exists which can be estimated visually to be about $\Delta\approx 6-7$ meV. The black dashed line highlights contributions of the spin wave scattering from MnO impurities that can be confirmed from MnO spin wave calculations shown in the Appendix.}
\label{fig4}
\end{figure}

Figure \ref{fig4} shows the low-energy INS data acquired with $E_{i}=$ 30 meV where it is clear that the spin wave spectrum possesses a gap, $\Delta$, at the G-type magnetic Bragg peaks.  This gap likely arises from single-ion anisotropy of the Mn ion. We note that the Mn ion in BaMn$_{2}$As$_{2}$ is sometimes discussed as having a formal valence of Mn$^{2+}$. However, Mn$^{2+}$ is a spin-only ($S=5/2$) ion with no magneto-crystalline anisotropy based on crystal-field effects.  Thus the presence of a spin gap (combined with the observation of a smaller ordered moment of 4 $\mu_{\mathrm{B}}$ instead of 5 $\mu_{\mathrm{B}}$ for $S=$5/2)  suggests that the Mn moment has an orbital component, perhaps more consistent with Mn$^{+}$ \cite{johnston11}.  The effect of the spin gap on hole doping is then interesting, because if the doped holes occupy Mn ions, then one would expect more Mn$^{2+}$ character and a closing of the spin gap with doping.

\begin{figure}
\includegraphics[width=0.9\linewidth]{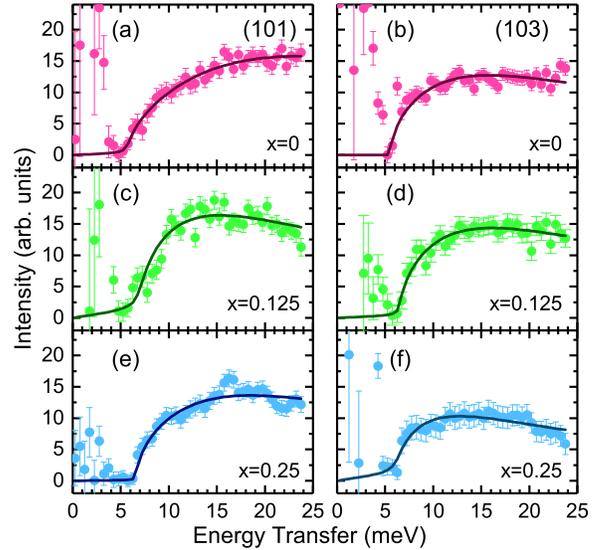}
\caption{\footnotesize  Energy cuts from $E_{i}=$ 74 meV data centered at $Q_{(101)}$ [left panels (a), (c), and (e)] and $Q_{(103)}$ [right panels (b), (d) and (f)] are used to determine the spin gap, $\Delta$, for different concentrations of Ba$_{1-x}$K$_{x}$Mn$_2$As$_2$ with $x=0$ [(a) and (b)], $x=0.125$ [(c) and (d)] and $x=0.25$ [(e) and (f)]. On each panel, the solid line is obtained by global fits of Eq. (\ref{LS}) to the data for each composition. The values of the fit parameters are given in Table \ref{tablo1}.}
\label{fig5}
\end{figure}

In order to determine the spin gap, we first performed constant-$Q$ energy cuts over a small range of $Q$ ($\pm$ 0.08 \AA$^{-1}$) centered at $Q_{(101)}$ and $Q_{(103)}$ for the $E_{i}= 74$ meV data. The background was estimated by similar cuts performed over the range from $Q=$ 1.75 to 1.9 \AA$^{-1}$ [in between (101) and (103)] and 2.35 to 2.4 \AA$^{-1}$ [above (103)] and subtracted from the (101) and (103) cuts after accounting for the $Q^{2}$ dependence, respectively. The resulting data are shown in Fig. \ref{fig5}. A sharp onset of magnetic intensity above $\sim$5 meV is observed for all compositions. The large errors in the intensity at the lowest energies arise from the subtraction of the strong elastic scattering contribution to the estimated background. 

The gap value for each composition is obtained by fitting the low energy spectrum to a damped simple harmonic oscillator expression
\begin{equation}
 S(Q,E)=\frac{AE^{'}\Gamma}{\Gamma^2+E^{'2}}            \label{LS}
\end{equation}
where $E^{'}$ is given by
\begin{equation}
 E^{'}=\frac{1}{\sqrt{2}}\Big[\Big((E^2-\Gamma_{\mathrm{s}}^{2}-\Delta^2)^{2}+4E^{2}\Gamma_{\mathrm{s}}^{2}\Big)^{1/2}+(E^2-\Gamma_{\mathrm{s}}^{2}-\Delta^2)\Big]^{1/2}.       \label{E}
\end{equation}
Here $A$ is the lineshape amplitude, $E$ is the energy transfer, $\Delta$ is the spin gap, $\Gamma_{\mathrm{s}}$ is a damping constant that describes the sharpness of the gap, and $\Gamma$ is the energy scale of the spin fluctuations.\cite{Lake}  Preliminary fits in which all parameters are allowed to vary independently generate similar values for $\Gamma$ and $\Delta$ from the (101) and (103) cuts. Thus, global fits to the (101) and (103) cuts for each K-substituted sample were performed to obtain $\Gamma$ and $\Delta$ whicle allowing the parameter $A$ to vary at each of the wavevectors, as shown in Table \ref{tablo1}.  

The results show that K-substitution has only a marginal effect on the magnitude of the spin gap.  Certainly, the gap is not closing with the addition of holes which is consistent with other observations that doped holes reside primarily in As $p$ bands, and therefore should not dramatically change the single-ion anisotropy of Mn ions.  $\Gamma$ appears to decrease with composition in these fits.  However, the physical interpretation of $\Gamma$ is complicated for several reasons.  Steep acoustic spin waves exist above the gap and changes in $\Gamma$ can arise from changes in the spin wave velocity, damping, or a combination of the two.  Also, the presence of MnO impurities in doped compositions can affect the determination of $\Gamma$.

\begin{table}
\caption {Energy scales and magnetic parameters of Ba$_{1-x}$K$_{x}$Mn$_2$As$_2$ as a function of composition. $k_\mathrm{B}T_\mathrm{N}$ is the actual N\'eel temperature measured by neutron diffraction.  $\langle E \rangle$ was determined numerically from Fig. \ref{fig3}.  Values of the spin gap ($\Delta$) and low energy damping ($\Gamma$) were determined from fits to Eq.~(\ref{LS}).  Other quantities (exchange constants and related parameters) are determined either directly or indirectly from fittings to the Heisenberg spin wave model as described in the text.  Errors are given as one standard deviation.}   
\label{tablo1}  
\begin{tabular}{ c | c | c | c  }
\hline\hline
Composition  ~             &~$x=0$~       &~$x=0.125$~        &~$x=0.25$ ~ \\
~Hole conc./Mn ~             &~$\delta=0$~       &~$\delta=0.0625$~        &~$\delta=0.125$ \\ \hline\hline
$k_\mathrm{B}T_\mathrm{N}$ (meV) \cite{Lamsal}  & 53.9(1) & 52.7 & 49.6(3)  \\
$\langle E \rangle$ (meV)		  & 66.0(3) & 64.7(4) & 59.2(3)  \\
$\Delta$ (meV)  &5.65(15)  & 6.71(19)   &6.28(21)  \\ 
$\Gamma$ (meV)  &18.1(8)  & 12.5(8)  &12.6(8)  \\ 
$SJ_1$ (meV)	  & 40.5(2.0)  & 41(4)   & 41(5) \\
$SJ_2$ (meV)	   &13.6(1.4)  & 13.9(2.0)    & 14.8(3.0)  \\
$SJ_c$ (meV)	 &1.8(3)    & 1.4(3) & 1.0(2)   \\  
$SD$ (meV)	 &  0.048(3)   & 0.068(7) &  0.060(8)\\
$J_{c}/J_{1}$	 & 0.044(8) & 0.034(8) & 0.024(6) \\
$J_{1}/2J_{2}$	 & 1.49(7) & 1.45 (10) & 1.38 (10) \\
$(k_\mathrm{B}T_\mathrm{N})_{\mathrm{MF}}$ (meV)  & 111(4) & 111(7) & 107(8) \\
$(k_\mathrm{B}T_\mathrm{N})_{\mathrm{MC}}$ (meV)  & 53.9(9) & 52.4(1.3) & 47.6(1.5)\\
$\hbar v_{ab}$ (meV \AA) & 196(3) & 195(6) & 181(6)\\  
$\hbar v_{c}$ (meV \AA) & 166(14) & 149(17) & 124(15)\\  \hline\hline

\end{tabular}\\ 
\end{table}
\subsection{Heisenberg model}
To ascertain more clearly the effect of hole doping on magnetic exchange interactions between Mn spins ($S_i$), the full ($Q$,$E$)-dependence of the G-type spin wave excitations were analyzed using the $J_{1}-J_{2}-J_{c}$ Heisenberg model. Here, $J_1$ and $J_2$ are the in-plane NN and NNN exchange interactions and $J_c$ is the out-of-plane NN AFM interaction. This Hamiltonian is written as 
\begin{equation}
\begin{split}
 H=J{_1}\sum_{\mathrm{NN}, ab}\textbf{S}_i \cdot \textbf{S}_j+ J{_2}\sum_{\mathrm{NNN}, ab}\textbf{S}_i \cdot \textbf{S}_j\\
+ J{_c}\sum_{\mathrm{NN}, c}\textbf{S}_i \cdot \textbf{S}_j+D\sum_i(S_i^z )^2           \label{hamiltonian}
 \end{split}
\end{equation}
The last term in Eq.~(\ref{hamiltonian}) represents the single-ion anisotropy where $D$ is the uniaxial anisotropy parameter appropriate for Mn spins directed along the $c$-axis.

In this article, as in reference \cite{johnston11}, the AFM interactions are represented by positive $J_{i}>0$. Within this model, the G-type ground state is only possible when $J_{1}$ and $J_{c}$ are AFM.  However, $J_{2}$ can be either FM or AFM. A ferromagnetic $J_{2}$ stabilizes the G-type order, whereas an AFM $J_{2}$ is a frustrating interaction that can destabilize G-type order.  When $J_{2}$ is AFM (as we find in the analysis below), classical G-type and stripe-type AFM ground state are possible and their energies are given by 
\begin{equation}
 E_{\mathrm{G}}=NS^2(-2J_1-J_c+2J_2)            \label{Eg}
 \end{equation}
 \begin{equation}
 E_{\mathrm{stripe}}=NS^2(-2J_2-J_c)         		 \label{Estripe}
\end{equation}
where $N$ is the total number of spins. The G-type state has a lower energy when $J_{1}>2J_{2}$. 

In order to simulate the full INS spectrum of $S(Q,E)$ for Ba$_{1-x}$K$_{x}$Mn${_2}$As$_{2}$, we use the linear spin wave approximation of the Heisenberg Hamiltonian given in Eq.~(\ref{hamiltonian}). The problem is cast in terms of the Holstein-Primakoff representation involving the boson spin operators. Using the Fourier transform of boson operators over the Mn sublattice, the following spin-wave dispersion relations are obtained for the G-type AFM structure in the $I4/mmm$ unit cell \cite{johnston11}
\begin{equation} \label{dispersion1}
\begin{split}
\bigg[\frac{\hbar\omega(\textbf{q})}{2S} \bigg]^2 = \big[2J_1 + J_c + D - J_2(2-\cos q_xa-\cos q_ya)\big]^2 \\
     - \Big\{ J_1\Big[ \cos\frac{(q_x+q_y)a}{2}+ \cos \frac{(q_x-q_y)a}{2} \Big]  + J_c\cos \bigg( \frac{q_zc}{2}\bigg) \Big\}^2 \\
 & 
\end{split}
\end{equation}
where $\textbf{q}$ is the wavevector measured relative to the G-type magnetic Bragg peak.

Various other quantities relevant for our discussion can be estimated from the Heisenberg spin wave theory.  Mean-field analysis shows that the N\'eel temperature is given by \cite{johnston11}
\begin{equation} \label{tn}
(k_\mathrm{B} T_\mathrm{N})_{\mathrm{MF}} = 2\big( 2J_1-2J_2+J_c \big)\frac{S(S+1)}{3}. \\
\end{equation}
This mean-field value of $T_{\mathrm{N}}$ will overestimate the actual transition temperature.  A more accurate relationship between $T_{\mathrm{N}}$ and the exchange constants can be obtained from fits to the ordering temperature obtained from classical Monte Carlo simulations of the $J_{1}-J_{2}-J_{c}$ Heisenberg model \cite{johnston11}.  This provides the following relation
\begin{equation} \label{tn_mc}
(k_\mathrm{B} T_\mathrm{N})_{\mathrm{MC}} = \frac{AJ_{1}S(S+1)}{B-\mathrm{ln}(J_{c}/J_{1})}\bigg[1-\bigg(\frac{J_{2}/J_{1}}{R_{1}}\bigg)^{R_{2}}\bigg] \\
\end{equation}
where $A(S+1)=$ 26.13, $B=$ 6.87, $R_{1}=$ 0.644, and $R_{2}=$ 1.082.  After obtaining the exchange constants from fitting the INS data as described below, very good agreement is found between $(k_\mathrm{B} T_{\mathrm{N}})_{\mathrm{MC}}$ and the measured value of $k_\mathrm{B} T_{\mathrm{N}}$, but the MF values are a factor of two too high, as shown in Table \ref{tablo1}.

For analysis of the low-energy spin excitations, one can also derive the value of the spin gap in terms of the single-ion anisotropy and exchange constants.
\begin{equation} \label{gap}
\Delta^2 \approx 8DS^2\big( 2J_1+J_c \big) \\
\end{equation}
The experimentally determined value of $\Delta \approx$ 6 meV will allow us to determine the anisotropy parameter, $D$, after the other exchange constants are determined by fitting (see Table \ref{tablo1}).  

For energies up to about 50 meV, the dispersion is roughly linear above the spin gap. The slope in the linear region is related to the spin wave velocity according to the relation
\begin{equation}
\hbar\omega(\textbf{q})=\big[\Delta^2 + \hbar^2 v_{ab}^2(q_x^2+q_y^2) + \hbar^2 v_c^2q_z^2  \big]^\frac{1}{2} ,          \label{dispersion2}
\end{equation}
where $\hbar\omega(\textbf{q})$ is the energy of the spin wave mode at wavevector $\textbf{q}$ close to the center of the first Brillouin zone. $v_{ab}$ and $v_c$ are the two unique spin wave velocities (in the $ab$-plane and along the $c$-direction, respectively) that are allowed based on the tetragonal symmetry of the crystal structure.

From Eq.~(\ref{dispersion1}), these velocities can be written in terms of magnetic exchange constants as
\begin{equation} \label{velocities}
\begin{split}
\hbar v_{ab}=2J_1Sa \Big[\Big(1-\frac{2J_2}{J_1} \Big) \Big(1+\frac{J_c}{2J_1} \Big)        \Big]^\frac{1}{2} \\
\\
  \hbar v_c=\sqrt{2}J_1Sc \Big[ \Big(\frac{J_c}{J_1}\Big) \Big(1+\frac{J_c}{2J_1}\Big)     \Big]^\frac{1}{2}\\
%
\end{split}
\end{equation}

In principle, the spin wave velocity can be estimated in a model-independent fashion by determining the slope from the raw INS data.  However, estimates of the spin wave velocity on powder samples are complicated by the averaging over different propagation directions and the sampling of reciprocal space.  As described below, since $v_{ab}$ is reasonably well defined from the data, it serves as an excellent fitting parameter for the Heisenberg model.

Solution of the linear spin wave equations-of-motion also produces the spin eigenvectors for each mode.  These eigenvectors can be used to calculate the magnetic neutron scattering intensity, $S_{\mathrm{mag}}(\textbf{Q},E)$, for each mode (where $\textbf{Q}=\mathbf{\tau}+\textbf{q}$ with $\mathbf{\tau}$ a magnetic Bragg peak position).  In the case of powder samples, the neutron scattering intensity must be averaged over all directions of \textbf{Q}.  This can be done numerically, providing an intensity function, $S_{\mathrm{mag}}(Q,E)$ that is suitable to compare directly to the background-subtracted magnetic scattering data.  A detailed explanation of these methods can be found elsewhere \cite{johnston11, Rob1}.

\subsection{Fits to the Heisenberg model}
The Heisenberg model spin wave calculations were performed for different sets of $SJ_1$, $SJ_2$ and $SJ_c$ values that were sampled by a Monte Carlo routine and compared with the INS data using a simple $\chi^2$ test. We performed evaluations of $\chi^2$ with two methods.  In the first method, we attempted to fit the full $(Q,E)$ data set [as shown in panels (a)--(c) of Fig.~\ref{fig2}] after making estimates of the background throughout.  In the second case, we summed the data over a range of angles, estimated and removed the energy dependence of the background and performed model fits to this energy spectrum [as shown in panels (a)-(c) of Fig.~\ref{fig3}]. Both methods gave similar results for the exchange constants.  However, the second method is preferred as it eliminates issues where background can dominate the evaluation of $\chi^2$ in regions where the magnetic scattering is absent.  

Here we focus on the results of the second method and discuss fits to the spectra in Fig. \ref{fig3}.  Powder-averaged calculations for a given set of $J_i$ were performed by Monte Carlo sampling 5000 $\textbf{Q}$-vectors on a given sphere of magnitude $Q$, giving the average energy-dependent neutron spectrum on each $Q$ sphere between $Q=$0 to 6.0 \AA$^{-1}$ with a step of $\Delta Q=$ 0.015 \AA$^{-1}$ (this amounts to the sampling of two-million different wavevectors for a given set of $J_i$).  The resulting powder-averaged magnetic scattering, $S_{\mathrm{mag}}(Q,E)$, was then averaged over the kinematically allowed $Q$ values for the experimental conditions with $E_i=144.7$ meV and $2\theta=$2.5--35 degrees (as shown in Fig.~\ref{fig2}). A reduced $\chi^2$ was evaluated by comparing these calculations to the data over the energy range from 30--110 meV after determining an overall scale factor that matched the integrated areas of the experimental and calculated spectra.   As many as 35000 different sets of $J_{i}$ were evaluated for each composition with finer steps in parameter space taken close to minima in $\chi^2$.  In general, fits that constrained the $J_i$'s to fixed values of $v_{ab}$ produced sharper minima in $\chi^2$.  Error bars in all fitted and derived parameters are reported to one standard deviation.  The errors of a given parameter are determined by its extremal values when the reduced $\chi^{2}$ is projected onto the parameter axis over the interval from $\chi^{2}_{\mathrm{min}}$ to $\chi^{2}_{\mathrm{min}}+3.53$ \cite{Press}.

This method produced reduced $\chi^{2}_{\mathrm{min}}$ values of 1.8, 6.2 and 12.0 for $x=0$, 0.125 and 0.25, respectively,  and the corresponding values of the exchange parameters at $\chi^{2}_{\mathrm{min}}$ are listed in Table \ref{tablo1}.  Calculations of the neutron spectra with these parameters, shown in Figs.~\ref{fig2} (d)--(f) and as the lines in Fig.~\ref{fig3}, show reasonable agreement with the data. For BaMn$_2$As$_2$, Figs.~\ref{fig2} and \ref{fig3} show that the Heisenberg model does an excellent job of representing the data.  However, the Heisenberg model fits become progressively worse with increased K substitution.  Part of the reason for the poorer fits in substituted samples comes from the higher prevalence of impurity phases that make estimations of the background more difficult. Although the poorer fits might also indicate that other interactions that are neglected in the analysis, such as damping, chemical disorder or longer-range interactions, may be present in the metallic samples.

For the parent compound, large values for $SJ_{1}$ and $SJ_{2}$ are found, consistent with high $T_{\mathrm{N}}$.\cite{Lamsal}  A relatively small interlayer exchange ($J_{c}/J_{1}\approx 0.05$) indicates that the magnetism is quasi-two-dimensional in nature.  A large AFM NNN interaction (with $J_{1}/2J_{2}\approx1.5$) highlights that some degree of magnetic frustration is present in BaMn$_{2}$As$_{2}$.  

The exchange values obtained for the parent compound are generally larger than the previous INS results described in Ref.~\cite{johnston11}, although derived quantities such as $v_{ab}$ and $J_{1}/2J_{2}$ show better agreement.  This study is likely to be an improvement over the previous work, since the current data are of much better quality and more robust fitting methods have been used.  

Generally speaking, relative errors on individual $J_{i}$ range from 5-20\% whereas combined parameters, such as $v_{ab}$, are much more narrowly defined with relative errors between 1--3\%.  Essentially, the fits to the INS data are most sensitive to combinations of the exchange parameters due to a high degree of correlation between the individual pairwise exchange constants.  Thus, while $SJ_{1}$ and $SJ_{2}$ do not change with composition within error, $SJ_{c}$, $v_{ab}$ and $v_{c}$ are all reduced by K substitution.

\begin{figure}
\includegraphics[width=0.9\linewidth]{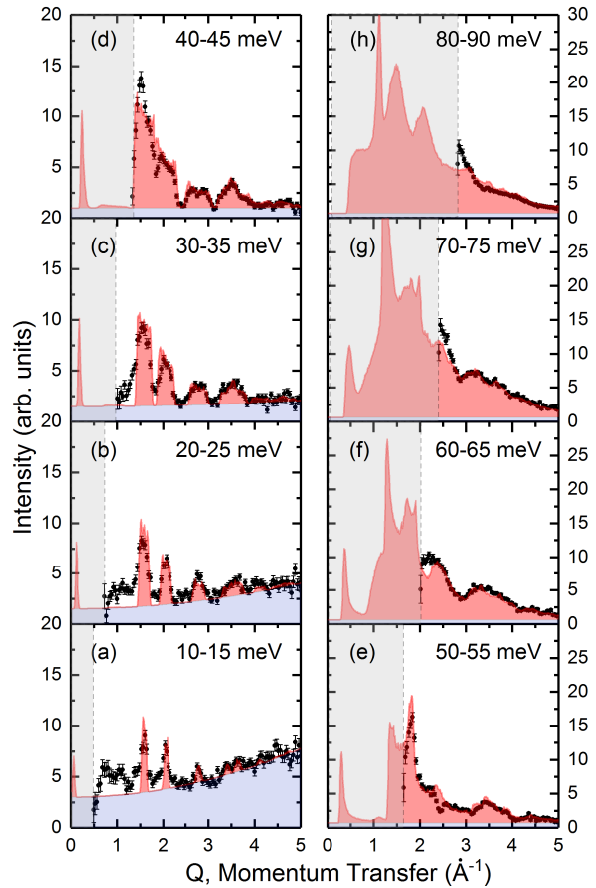}
\caption{\footnotesize  Different constant-energy $Q$-cuts of the BaMn$_{2}$As$_{2}$ data at $E_{i}=$144.7 meV as a function of momentum transfer (black dots) averaged from (a) 10-15 meV, (b) 20-25 meV, (c) 30-35 meV, (d) 40-45 meV, (e) 50-55 meV, (f) 60-65 meV, (g) 70-75 meV and (h) 80-90 meV.  The red line and red shaded area correspond to Heisenberg spin wave fits with parameters in Table \ref{tablo1}.  The background (blue shaded area) is estimated by a $Q$-independent term plus a quadratic term representing the phonon contribution.  The gray shaded rectangle on the left represents momentum transfers that are inaccessible kinematically at $E_{i}=$144.7 meV.}
\label{fig6}
\end{figure}

A more quantitative picture of the agreement between the model and data can be ascertained from a series of constant-energy $Q$-cuts, as shown for BaMn$_{2}$As$_{2}$ in Fig.~\ref{fig6}. In these plots, the model calculations (red) are added to an estimate of a quadratic plus constant background contribution (blue).  Below 40 meV, the constant-energy $Q$ cuts show a series of peaks in the linear acoustic spin wave regime.  The higher-energy cuts correspond to spin waves close to the Brillouin zone boundary.  The spin wave dispersion flattens out in these regions of reciprocal space, leading to large contributions to the powder-averaged spin wave scattering intensity and a more  continuous $Q$ dependence.  The spin wave model calculations capture the position, intensity and $Q$-width of all features quite well, despite the presence of phonon scattering and other background features of the data.  Certain features of the data are not captured by our model calculations and simple background model.  For example, additional scattering is observed at lower $Q$ close to the direct beam at most energy transfers suggesting additional $Q$-dependent background contributions.

\begin{figure}
\includegraphics[width=0.9\linewidth]{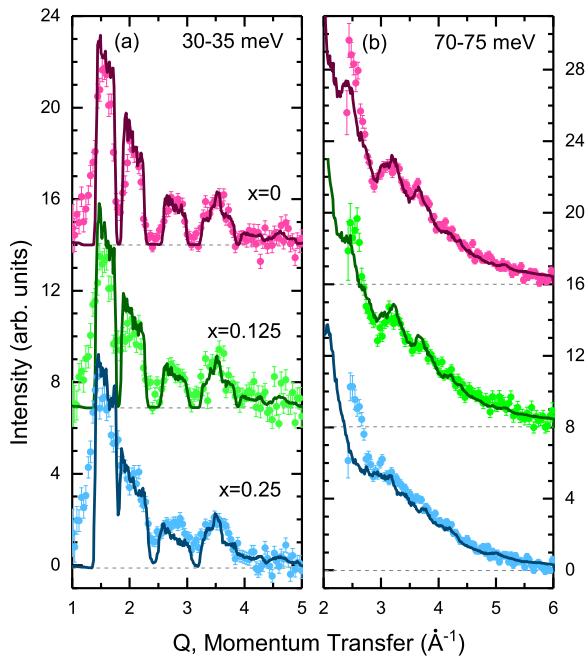}
\caption{\footnotesize  A comparison of constant-energy cuts for different compositions of Ba$_{1-x}$K$_{x}$Mn$_{2}$As$_{2}$ at (a) 30-35 meV and (b) 70-75 meV.  Dots correspond to experimental data with the estimated background subtracted. Lines correspond to Heisenberg model calculations of the spin waves.  Plots are offset vertically by 7 units in panel (a) and 8 units in panel (b).}
\label{fig7}
\end{figure}

The effect of K substitution on the constant-energy $Q$-cuts is shown in Fig.~\ref{fig7}.  The main effect is a broadening of the acoustic spin wave peaks most easily observed as broadening of the (101) and (103) acoustic spin wave features in Fig.~\ref{fig7} (a) from 30-35 meV.  It might be tempting to assume that the spin wave velocity $v_{ab}$, which mainly contributes to the powder-averaged dispersive features at (101) and (103), is strongly reduced with K substitution. However, Table \ref{tablo1} shows that the fitted values of $v_{ab}$ are reduced only 8\% from $x=0$  to $x=0.25$.  Rather, the model fits indicate that this broadening is caused mainly by the reduction of $SJ_{c}$. While this reduces $v_{c}$ by 33\%, the main effect of reducing $SJ_{c}$  on the (101) and (103) dispersive features is to smear out the intensity of the spin wave cone in $Q$ due to powder averaging.  This highlights the care that must be taken in interpreting the slope of the dispersive features in powder data, since it is affected by both the velocities and the dimensionality of the magnetic interactions. At higher energies where the spectra are more continuous, the effect of substitution on the $Q$-dependence is more subtle.  In all, the models capture both the energy and momentum dependence of the spin wave scattering for all three compositions and small differences between the spectra are accounted for by minor systematic changes in the exchange constants.

\section{Conclusion}
In summary, AFM spin waves of BaMn$_{2}$As$_{2}$ and K-doped Ba$_{1-x}$K$_{x}$Mn$_{2}$As$_{2}$ powders with $x=$0.125 and 0.25 were measured using INS. We conclude that the hole doping and the resulting metal-insulator transition only weakly affect the spin waves, even at hole concentrations up to $\delta = x/2 =$ 12.5\%.  This might be expected from the simple fact that $T_\mathrm{N}$ decreases by less than 10\% between the parent and $x=0.25$ compounds. Using Eq. (\ref{tn_mc}), the observed decrease in $T_{\mathrm{N}}$ with doping has almost equal contributions from an increase in quasi-two-dimensionality (a decrease in $J_{c}/J_{1}$ from 4.4\% down to 2.4\%) and an increase in the in-plane magnetic frustration (a decrease in $J_{1}/2J_{2}$ from 1.49 to 1.37).

In the analysis of the diffraction data by \cite{Lamsal}, it is also found that the Mn magnetic moment shows no large changes with doping.  This is consistent with the observation that the spin gap shows little or no change with composition.  The spin gap is associated with single-ion anisotropy of the Mn ion, and this result (in combination with the stability of the magnetic moment) allows us to conclude that hole doping does not appreciably change the orbital occupancies of the Mn ion. 

Taken together, these observations are consistent with doped holes residing primarily in As $p$ bands where they do not directly affect the Mn ion's net magnetic moment or orbital occupancy, but can influence the magnetic exchange interactions.  In particular, $J_{c}$ and $J_{2}$ are likely to have significant contributions from superexchange processes mediated through bridging As ions.  Any reduction of the electron density on the As sublattice through hole doping would affect these superexchange interactions.

Finally, we discuss the interesting observation that FM coexists with AFM order in K and Rb doped BaMn$_2$As$_2$ with compositions beyond $x\approx 0.16$.  The origin of this FM component has been proposed to arise from canting of the Mn spins \cite{Lamsal, mazin}, or from As band ferromagnetism \cite{griffin}.  Recent XMCD measurements support a picture where FM arises from doped holes on the As sublattice, which is at least consistent with our observations here. To go further, it should be possible to observe FM fluctuations directly in our INS experiments and address this controversy.  However, we see no evidence of any FM signal in our $x=$ 0.25 sample where FM order is expected below $\sim$100 K.  The small size of the ordered FM moment ($<$ 0.2 $\mu_{\mathrm{B}}$) would make it very difficult to observe a FM signal, especially in powder samples.  Also, the presence of MnO magnetic impurities in our $x=0.25$ sample further hampers any search for a definitive signal from FM fluctuations at energies below 25 meV.  Additional INS measurements on single-crystal samples with higher K or Rb compositions are necessary to explore this point further. 

\section{Acknowledgements} 
MR would like to thank N. Ozdemir for her support.  Work at the Ames Laboratory was supported by the Department of Energy, Basic Energy Sciences, Division of Materials Sciences and Engineering, under Contract No. DE-AC02-07CH11358. This research used resources at the Spallation Neutron Source, a DOE Office of Science User Facility operated by the Oak Ridge National Laboratory.  Work at ITU is supported by TUBITAK 2232.
\\

\section{Appendix}
\subsection{Background estimates}

The powder-averaged inelastic neutron scattering intensity is comprised of several contributions from magnetic, $S_{\mathrm{mag}}(Q,E)$, and nuclear (phonon) neutron scattering, $S_{\mathrm{ph}}(Q,E)$, in addition to scattering from the empty can, $C(Q,E)$.  Other instrumental and background contributions are combined into a general background function, $B(Q,E)$.  Ignoring other experimental complications, such as sample absorption, we approach the analysis by approximating the total cross-section with the equation below.
\begin{equation} \label{total}
I(Q,E) = S_{\mathrm{mag}}(Q,E) + S_{\mathrm{ph}}(Q,E) + B(Q,E) + C(Q,E) \\
\end{equation}

The general prescription for analysis of the magnetic scattering is to first subtract off an independent measurement of the empty can, which is performed in all instances.  The next task is to estimate both $S_{\mathrm{ph}}$ and $B$ and subtract them as well.  In general, both $S_{\mathrm{ph}}$ and $B$ can be fairly complicated functions of both $Q$ and $E$ and it is difficult to determine the full functional dependence, especially when impurity phases are present in the samples.   We present the results for obtaining the energy spectra (as in Fig.~\ref{fig3}) and also the constant-energy $Q$-cuts (as in Fig.~\ref{fig6}).  

For the purpose of determining the magnetic energy spectra in Fig.~\ref{fig3}, the total background $S_{ph}+B$ is estimated from suitably scaled high-angle data where phonon scattering is strongest and magnetic scattering is absent.
\begin{equation} \label{smag}
S_{\mathrm{mag}}(E) = S_{\mathrm{low}}(E)- KS_{\mathrm{high}}(E) \\
\end{equation}

\begin{figure}
\includegraphics[width=0.7\linewidth]{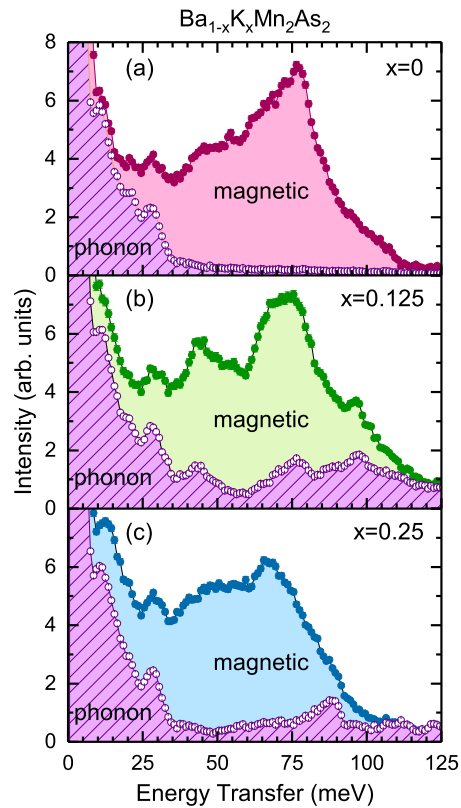}
\caption{\footnotesize  High-angle spectra (empty symbols) and scaled low-angle spectra (solid symbols) at $E_{i}=$144.7 meV for (a) $x=0$, (b) 0.125, and (c) 0.25.  The shaded region corresponds to the estimated magnetic scattering spectrum $S_{\mathrm{mag}}(E)$ and the hatched shaded area is $KS_{\mathrm{high}}(E) \approx S_{\mathrm{ph}}(E)+B(E)$. }
\label{figa1}
\end{figure}

For the $E_{i}=$ 144.7 meV data, the high-angle spectrum $S_{\mathrm{high}}(E)$ averaged from 42.5--83.5 degrees is scaled by $K$ and subtracted from the low-angle spectrum $S_{\mathrm{low}}(E)$ averaged from 2.5--35 degrees. It is clear from Fig.~\ref{figa1}(a) that the $x=0$ sample is relatively clean from impurities as there are no phonon features found above 40 meV.  The scale factor $K$ is chosen by eye to completely eliminate obvious phonon peaks for $x=0$ and was found to be $K=$ 0.35.  On the other hand, the $x=$0.125 and 0.25 samples have additional phonon peaks above 40 meV due to impurity phases in the powder sample.  Whereas the low energy phonons ($<$ 50 meV) scale with the same $K$ factor, a different scale factor of 0.6 and 1.4 for the $x=$0.125 an 0.25 samples, respectively, is employed for the higher energy phonons ($>$ 50 meV) that originate from sample impurities. The low-angle and scaled high-angle spectra are shown in Fig.~\ref{figa1}.  The difference, $S_{\mathrm{mag}}$(E) is shown as the shaded region in Fig.~\ref{figa1} and also as the data in Fig.~\ref{fig3}.

For the constant-energy $Q$ cuts shown in Fig.~\ref{fig6}, a simple form is chosen to represent the $Q$-dependent background for each independent cut at an average energy transfer of $E$.  This form is simply a quadratic term intended to represent phonon scattering plus a constant term, $S^{E}_{\mathrm{ph}}(Q)+B^{E}(Q)=U Q^{2}+F$.  The parameters $U$ and $F$ are given in Table \ref{tablea1} were chosen for each $Q$ cut in Figs.~\ref{fig6} and \ref{fig7} as a guide to the eye.

\begin{table}
\caption {Parameters ($U$, $F$) employed in the background estimates for constant-energy $Q$ cuts found throughout the paper.}   
\label{tablea1}  
\begin{tabular}{ c | c | c | c  }
\hline\hline
Energy range  ~             &~$x=0$~       &~$x=0.125$~        &~$x=0.25$ ~ \\ \hline\hline
10-15 meV  & (3, 0.19) & - & -  \\
20-25 meV  & (1.5, 0.1) & - & -  \\
30-35 meV  & (1.5, 0.02)  &  (1.6, 0.08)  & (1.3, 0.04)  \\ 
40-45 meV  & (1, 0)  & -  &-  \\ 
50-55 meV  & (0.7, 0)  & -   & -\\
60-65 meV  & (0.6, 0)  & -    & - \\
70-75 meV  & (0.6, 0)    & (1.3, 0.04) & (0.7, 0)   \\  
80-90 meV  & (0.7, 0) & - & -  \\
 \hline\hline
 
\end{tabular}\\ 
\end{table}

 \subsection{Spin wave scattering from MnO impurities}

\begin{figure}
\includegraphics[width=0.7\linewidth]{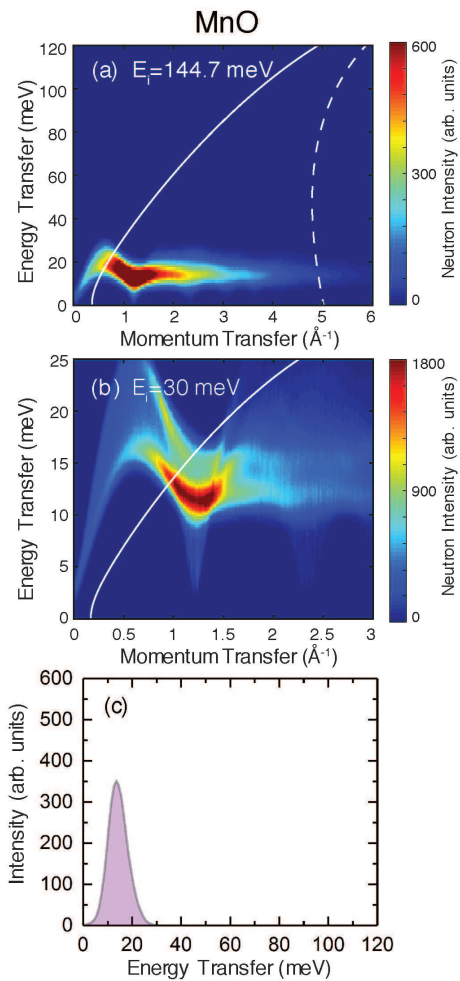}
\caption{\footnotesize  Calculations of the powder-averaged spin wave scattering from MnO.  The full neutron intensity $S(Q,E)$ is shown for an incident energy of (a) 144.7 meV and (b) 30 meV.  The solid line denotes the kinematic limit near forward scattering ($2\theta=2.5$ degrees) and the dashed line in panel (a) corresponds to a scattering angle of $2\theta=35$ degrees.  Panel (c) shows the MnO spin wave spectrum obtained at $E_{i}=$ 144.7 meV after averaging over the angle range in (a).}
\label{figmno}
\end{figure}

MnO impurities were discovered in our doped samples, most notably in the $x=0.25$ sample.  MnO is an antiferromagnet at the temperatures studied and clear spin wave scattering is observed below 25 meV from MnO impurities.  MnO spin waves have been studied in detail by inelastic neutron scattering and the data have been fit with linear spin wave theory. Using the exchange parameters determined in Ref.~\cite{kohgi}, we calculated the expected scattering from polycrystalline MnO impurities with $E_{i}=$ 144.7 meV, as shown in Fig.~\ref{figmno}.  Figure~\ref{figmno}(a) shows that the spin waves contribute strongly below 25 meV and display dispersive features up to about 3 \AA$^{-1}$.  The angle-averaged spectrum is shown in Fig.~\ref{figmno}(b) and can be compared to Fig.~\ref{fig3}(c), where the peak corresponding to MnO spin wave scattering is apparent in the data.

\subsection{Uniqueness of fitted exchange values}

\begin{figure}
\includegraphics[width=0.9\linewidth]{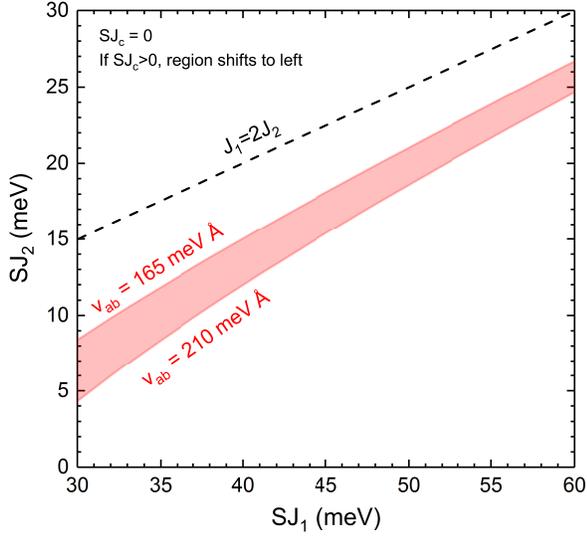}
\caption{\footnotesize  Region of $SJ_{1}-SJ_{2}$ parameter space delineated by boundaries of $v_{ab}$ from 165 to 210 meV \AA\ with $SJ_{c}=0$ is shown as the shaded region.  For a G-type ground state, values of $(SJ_{1},SJ_{2})$ above the dashed line correspond to the condition $J_{1}<2J_{2}$ and are not allowed.}
\label{figpspace}
\end{figure}

Modeling using the spin wave approximation to the Heisenberg model was performed against a set of exchange parameters $SJ_1$,$SJ_2$ and $SJ_c$ as explained in the main text. We discovered that $SJ_{1}$ and $SJ_{2}$ are highly correlated and different combinations of these parameters (such as $v_{ab}$) result in more sharply defined minima in $\chi^{2}$.  This is mainly due to the fact that $v_{ab}$ is reasonably well determined by the slope of the acoustic spin waves at low energies with a value of roughly 200 meV \AA.   A reasonable range of $v_{ab}$ from 165-210 meV \AA\ was chosen for fitting, which dramatically reduces the necessary $SJ_{1} - SJ_{2}$ parameter space that needs to be explored, as shown in  Fig.~\ref{figpspace}.

\begin{figure}
\includegraphics[width=1\linewidth]{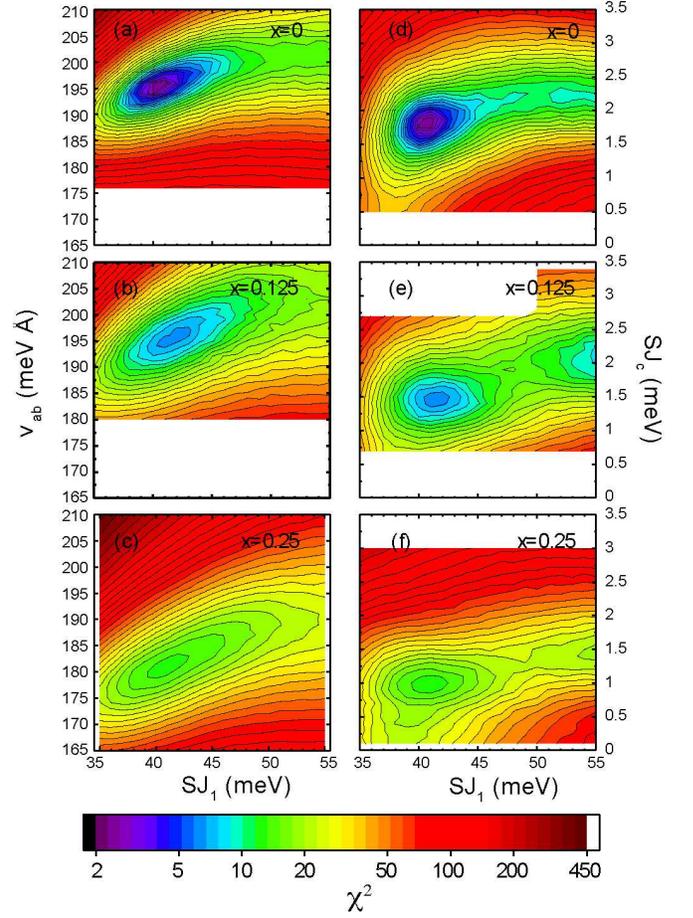}
\caption{\footnotesize  Goodness of the fit, $\chi^{2}$, plotted in the $SJ_{1}$ - $v_{ab}$ plane for fixed $SJ_{c}$ [panels (a) - (c)] and the $SJ_{1}$ - $SJ_{c}$ plane for fixed $v_{ab}$ [panels (d) - (f)].  Plots are shown for each composition $x=0$ [panels (a) and (b)], $x=0.125$ [panels (c) and (d)], and $x=0.25$ [panels (e) and (f)]. Minima in $\chi^{2}$ are apparent for the cuts shown.  In panels (a) - (c), the parameter $SJ_{c}$ is held fixed at 1.8, 1.4, and 1.0 meV, respectively.  In panels (d)-(f), the parameter $v_{ab}$ is held fixed at 196, 195 and 181 meV \AA, respectively.  The value of $\chi^{2}$ is indicated by the logarithmic color scale.}
\label{figchisq}
\end{figure}

Therefore, we chose three independent parameters, $SJ_1$,$v_{ab}$ and $SJ_c$ as the main parameter set for fitting.  The dependence of $\chi^{2}$ on these parameters is shown for different cuts in this three-dimensional parameter space when either $SJ_{c}$ [Fig.~\ref{figchisq} (a) - (c)] or $v_{ab}$ [Fig.~\ref{figchisq} (d) - (f)] is held fixed at the value that minimizes $\chi^{2}$.  For $x=0$, a strong and deep local minimum exists in $\chi^{2}$.  For $x=$0.125 and 0.25, the minimum value of $\chi^{2}$ is higher, and therefore the goodness of fit is worse.  In addition, another local minimum in $\chi^{2}$ develops at higher values of $SJ_{1}$ and $SJ_{c}$ [see Fig.~\ref{figchisq} (e)]. However, we do not expect the parameters to change too drastically, so we report our best parameters relative to the local minimum found in the $x=0$ data.  In addition, parameters in the second minima have values of $(k_\mathrm{B} T_{\mathrm{N}})_{\mathrm{MC}}$ (Eq.~\ref{tn_mc}) that do not conform to the observed N\'eel temperature.

\end{document}